# Music Recommendation on Spotify using Deep Learning

Chhavi Maheshwari[a]*

*[a]Manipal University Jaipur, Jaipur, Rajasthan, India*

**Abstract**

Hosting about 50 million songs and 4 billion playlists, there is an enormous amount of data generated at Spotify every single day - upwards of 600 gigabytes of data (harvard.edu). Since the algorithms that Spotify uses in recommendation systems is proprietary and confidential, code for big data analytics and recommendation can only be speculated. However, it is widely theorized that Spotify uses two main strategies to target users' playlists and personalized mixes that are infamous for their retention – exploration and exploitation (kaggle.com). This paper aims to appropriate the filtering using the approach of deep learning for maximum user likeability. The architecture achieves 98.57% and 80% training and validation accuracy respectively.

## 1. Introduction

The entertainment sector has seen formative changes recently due to big data and its analysis. Major players in each major category - OTT, audio, and video have utilized artificial intelligence for personalization in a targeted demographic. What is arguably considered the best music and podcast service platform due to ease of access, contributions to ending music piracy, and most of all, tailored playlists, Spotify is now the biggest audio player with about 35% of the market share. Spotify has perfected their recommendation system, which has attributed to their customer retention all along. This recommendation system has been the basis for many features – from song radios to personalized mix playlists to smart shuffle, i.e. song recommendations for a playlist.

The current recommendation strategy uses many machine learning algorithms to personalize content for its users. This involves analyzing various data points about the user, for example - listening history, search history, playlist creation, and data on other behavior to create a personalized music experience.

There are three alleged major components to this recommendation engine: Content Filtering (Explorative and Exploitative), Natural Language Processing, and Audio Models. Content filtering is the data analysis of listening habits of millions of users to identify patterns and similarities in their music tastes. This also involves recommending songs that are played by users with a similar taste. Natural language processing analyzes text such as playlist titles, tags, and searches to predict their preferences. Audio models analyze the intricate acoustic properties of songs, hence analyzing the song itself.
Other personalized playlists like Discover Weekly provide users with a hitherto new combination of carefully selected music. Another reason the platform is beloved is the curated playlists. These include genre, mood, and activity, as well as personalized radio stations based on a particular song, artist, or genre.

---

* Corresponding author: *E-mail address:* chhavi30m@gmail.com

## 1.1. Historical Context

The advent of music platforms (hosting and sharing) started in the 1990s and 2000s with groundbreaking applications like Napster and iTunes. These revolutionized the way music was and is consumed. Pandora Radio, launched in 2005 was the first platform that introduced personalized recommendations in the form of "song radios". In 2006, Spotify was founded in Sweden and started offering users access to a vast music catalog. It also introduced the freemium model, which has since become the dominant monetization strategy in the industry. Since then, Spotify has established a near-monopoly position in the market. Its user retention is often credited to its highly effective recommendation systems. Building on this success, later, Spotify expanded its audience by acquiring the podcast space, hosting some of the most popular podcasts exclusively on its platform. This strategic focus on personalization and data-driven approaches, including data analytics, machine learning, and deep learning, has been instrumental in Spotify's remarkable success.

## 1.2. Application of Deep Learning

Deep Learning is crucial in Spotify's working. It helps identify the most important features from user datapoints, conduct audio analysis to give explorative recommendations, process natural language in search bars, ad targeting based on various factors like demography and maintaining content quality. Deep learning has made significant contributions to music recommendation personalization. With its ability to analyze large amounts of data and extract complex patterns, deep learning models can effectively capture the intricate relationships between users' music preferences and various factors such as genres, moods, and listening habits.

A key advantage of deep learning in music recommendation is the capability to learn high-level representations of music features. Deep learning models can extract important information, capturing both low-level acoustic and high-level semantic features using various convolutional layers. This helps the models understand many underlying characteristics and make "good" preferential recommendations. Some deep learning models also leverage time-sequential data like users' listening histories to capture time, season, and phase dependencies and thus, make personalized recommendations. Recurrent Neural Networks (RNNs) and its variants like Long Short-Term Memory (LSTM) networks are used in music recommendation systems to model this sequential nature of user behavior. For example, after listening to Rolling Stones, a user might listen to Led Zeppelin, Aerosmith, and The Beatles. If the data is sequential, a logical conclusion can be drawn that the user enjoys rock band music. Methodologies like these effectively capture the dynamics of user's preferences over time and adapt recommendations accordingly.

For other supporting tendencies, deep learning models can incorporate various sources of data beyond just musical audio features. They can incorporate textual data from song descriptions, user reviews, or social media posts to gain a deeper understanding of the content and context of songs. This allows for fine-grained personalization based on users' preferences and interests. Continuing with the previous example, searching for Led Zeppelin or "All time rock hits" can also support the hypothesis that the user likes rock.

## 1.3. Exploitative and Explorative Content Filtering

Essentially, Spotify allegedly uses a combination of two algorithms to provide user-personalized recommendations: exploitative and explorative content. Exploitative recommendations make use of existing data on likes and dislikes (Schildt et al., 2005). Explorative recommendations, on the other hand, make keen use of audio analysis to determine what a user may like that might be out of their "comfort zone". For example, if a user is mainly pop music listener, their daily personalized playlists will usually consist of the same. These are exploitative recommendations. However, every now and then, a pop-punk song will be introduced in the playlist to expand the taste and gauge how the user reacts to the song. This will be further used to update recommendations. This was an explorative recommendation. Exploitative content usually requires previous data to work from, whereas explorative content uses the feature data from the song itself. The former is a classic example of network analysis, whereas the latter is a content-based exploration.

*1.3.1. Exploitative Filtering:*

Exploitative filtering or collaborative filtering makes use of two methods - history-based recommendations and socially similar recommendations. History-based recommendations are based on what the user themself has listened to and shown an interest in before. Listening history is a crucial element here. Socially similar recommendations, on the other hand, make use of a "network". This network makes use of neighbors that are users with most similar, similar, and dissimilar tastes. This recommends 'User A' music that other close neighbors have watched (Fig. 1). This similarity between User A (active user) and another user is calculated using Pearson coefficients (Sedgwick, 2012).

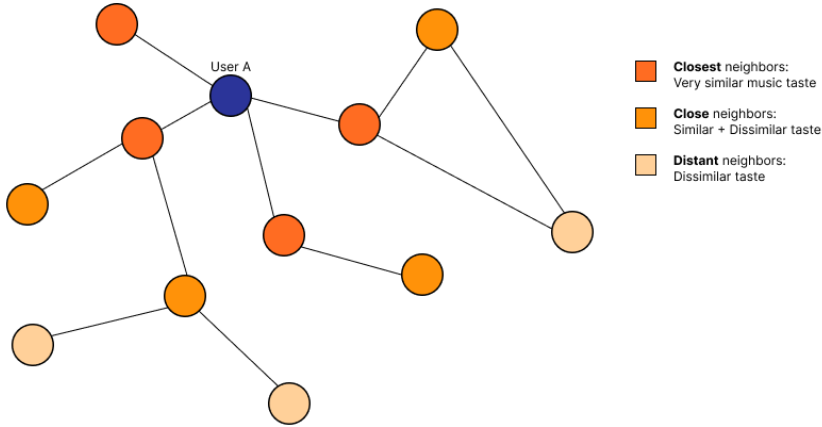

Figure 1: Exploitative filtering based on social similarity

$$c_{a,u} = \frac{covar(r_a, r_u)}{\sigma_{r_a} \sigma_{r_u}} \quad (1)$$

Here, $covar(r_a, r_u)$ represents covariance between ratings of active user (a) and another user (u). $\sigma_{r_a}$ and $\sigma_{r_u}$ represent the standard deviations in the ratings of active and another user respectively. There are, however, many obstacles in exploitative filtering that require the usage of another algorithm in conjunction for effective recommendation. One, there is a requirement for substantial data to start off of, so that recommendations can be made, which can obviously not be the case every time. Two, even if there is a rating matrix available for the active user, there is heterogeneity in the overall taste because of how each user consumes content. Three, popularity bias makes data skewed. Particularly, given the state of the music landscape now, popular songs may be recommended to users more simply because of the frequency of occurrences. However, this does not concur for users with different and arguably unique taste.

*1.3.2. Explorative Filtering:*

Explorative filtering solves the problems that exploitative comes up with. The recommendations are based on the sole characteristics of the content (Pazzani & Billsus, 2007). These include factors like tempo, acousticness, energy and many other manual tags. A correlation between those values helps give an insight on the user likeability patterns.

## 2. Related Work

*2.1. Literature Review*

There have been many speculations regarding the exact hierarchy of algorithms that Spotify uses (Gulmatico et al., 2022). Kaminski and Ricci aimed to predict if 13 audio factors could determine the success of a song (Kaminski et al., 2009). Many papers use song attributes and analyze to attempt and build the recommendation backend (Allawadi & Vij, 2023).

4However, traditional research remains rather minimal on such a commercial algorithm. The attempts to build a stable recommendation system on available data are hosted on Kaggle. The techniques range from supervised to semi-supervised learning, using algorithms like logistic regression, naive bayes, k-nearest neighbors, support vector machines etcetera.

Deep learning on the other hand can manage to capture more intricate patterns in data. It has better learning ability for hierarchical representations of data and complex patterns. This can also scale easily for larger amounts of data. It also performs better with non-linearity.

*2.2. Motivation*

With a year-on-year increase in revenue and more than 500 million monthly users, Spotify has clearly mastered personalization. Near-replications of the recommendation algorithms include mostly machine learning approaches. However deep learning can encapsulate more intricate and complex patterns between various features of data.

Some other advantages of using deep learning can be:

- Better performance: Capturing and predicting convoluted datapoints
- Flexibility: Various representations to predict various outcomes using the same data
- Continual learning: Allows for continual learning and adapting to the new data

# 3. Proposed Work

*3.1. Dataset Loading*

The dataset had to have various sonic features like danceability, tempo, energy, key, loudness, acousticness etc. Thus, this was procured from Kaggle. The following features were used to determine recommendations:

- Danceability: Measure of how likely one is to dance to the sound.
- Energy: Measure of the dynamics of energy of the sound
- Loudness: Decibel measure of the sound
- Energy: Measure of the dynamics of energy of the sound
- Acousticness: 0 to 1 measure of whether the track is acoustic
- Instrumentalness: Measure of how likely the track is to have no spoken word vocals
- Liveness: Reverberation and ambiance measure of the track
- Valence: Measure of the musical positiveness conveyed by the track
- Tempo: Measure of speed of the track, measured in bpm
- Time signature: measure of number of counts in each measure and types of notes that receive one count

In addition to the multiple features, "History" is whether the user liked the song or not. Then to get a better viewpoint of the data's relation with likeability, graphs between the attributes and target variable were plotted (Fig. 2).

A correlation matrix was used to encapsulate the entire relationship. A correlation matrix is a N×N table of statistical measures that indicate the degree of relatedness of two variables. It is commonly used in data analysis to identify how increase and decrease in factors affects others. This is done using a range of -1 to 1, where -1 indicates a perfect absence of correlation and 1 indicates perfect correlation between two variables. (Fig. 3).

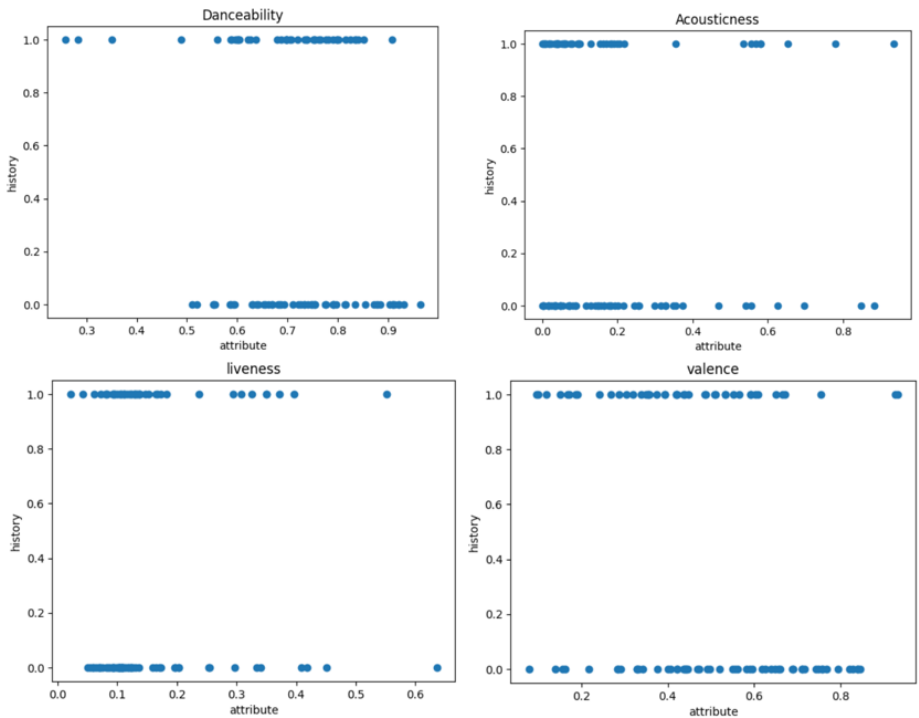

Figure 1: Quantitative relation of different auditory factors to likeability history

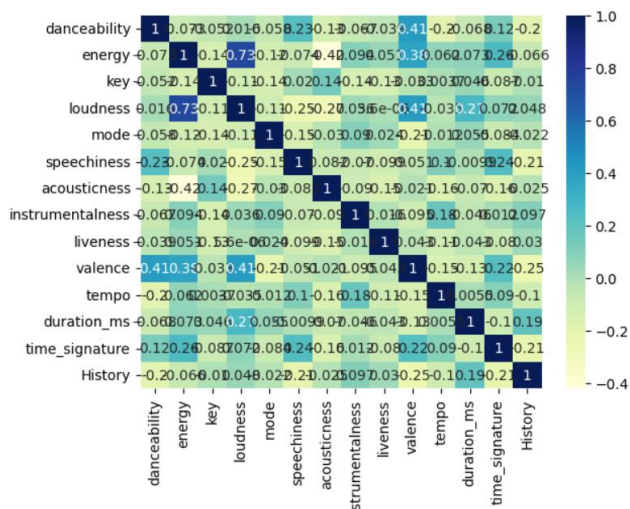

Figure 2: Correlation matrix of all attributes

*3.2. Preprocessing*

Data preprocessing is a crucial step in any data analysis or artificial intelligence approach. It transforms raw and unstructured data into a more suitable format for analysis. Data preprocessing contributes to better and viable results in many ways:

- Data quality: It improves the quality of the data by identifying and correcting errors, missing values, and data inconsistencies. This ensures that the data is reliable and modellable, and that the further steps can be carried out. In this case, the original dataset was free of null values.
- Feature selection: Data preprocessing helps to identify the most relevant features or variables for the analysis or modeling. This can reduce the dimensionality of the data, making it easier to work with and improving the accuracy of the analysis or model. Here, the 13 features that denote sonic characteristics were selected and time-sequential data was left out. This was done to keep the focus on content filtering.
- Data normalization: Data can be normalized by scaling it to a range of 0 to 1, both inclusive. This reduces the impact of outliers and allows for smoother analyzing without loss of properties or patterns. Since characteristics like loudness have expansive scale, these can be highly contributive in changing the outcome. Thus, normalization here is necessary.
- Data transformation: Data preprocessing can transform the data by applying mathematical functions or statistical techniques. This can reveal hidden patterns or relationships in the data, making it easier to interpret and analyze.

For a better representation and more accurate predictions and reduce the extremities of the range in each attribute, min_max_scaler was employed for each of the 13 features. This helped scale each value to between 0 and 1.

$$F(x) = \frac{x - x_{min}}{x_{max} - x_{min}} \quad (2)$$

This helped achieve the following:
- Data standardization: By transforming to a uniform scale, the data is easier to analyze.
- Preservation of relationships: This transformation still preserves the relative order and feature vector distance between variables and their relationships.
- Enhancement of convergence: The standardization allows for lesser error rates on larger-scale features, thus allowing for faster and more efficient convergence.
- Reduced sensitivity to outliers: Due to this scaling, outliers in large-scale features do not tend to have as big an impact as they would.
- Lesser numerical instabilities: By bounding this scaling to a range, a small variation in input does not tend to adhere to the chaos theory.

*3.3. Architecture*

Keras is a deep learning framework written in Python, particularly used in deep learning models. It is designed to enable fast experimentation with pre-defined layers, models, and functions. It has a user-friendly API that makes it highly accessible to both beginners and experts. It has various configurations for CPU and GPU computations, making it efficient for training large neural networks.

In this architecture, the main challenge was capturing the numerical patterns while also simultaneously reducing spatial dimensions. This is achieved by multiple dense layers. A dense layer is a group of neurons that connect every neuron in the input layer to the ones in the output layer. It is also called a fully connected layer. This is the most fundamental layer in deep learning and is typically used between input and output layers. Here, linear transformation was performed by the layers by multiplying it with a weight matrix and adding a bias vector. The output is then passed through the specified activation functions, introducing non-linearity for better learning.

ReLU (Rectified Linear Unit) activation function was used for simplicity and effectiveness, with a final sigmoid function for binary classification - like and dislike.

$$f_{ReLU} = max(0, x) \quad (3)$$

$$f_{sigmoid} = \frac{1}{1+e^{-x}} \quad (4)$$

### 3.4. *Hyperparameter Tuning*

The optimizer function used was Adam. Adam makes use of individual adaptive learning rates for each parameter. It keeps a running estimate of the first and second moments of the gradients, which are used to adjust the learning rate for each parameter. This helps to improve the performance of the optimizer, especially when dealing with sparse gradients.

| **Algorithm 1. ADAM algorithm** |
| --- |
| • Compute the gradient of the loss function with respect to the weights. |
| • Compute the mean of the gradient (first moment or $m_0$) and the uncentered variance of the gradient (second moment or $m_1$). |
| • Update the weights using the following formula: $w = w - \alpha * (m_0/\sqrt{m_1 + \epsilon})$ |

## 4. Results and Discussion

### *4.1. Results*

#### *4.1.1. Accuracy*

The accuracy is then computed as the fraction of true predictions in total predictions. It is calculated using the formula:

$$Accuracy = \frac{TN + TP}{TN + TP + FP + FN} * 100 \quad (5)$$

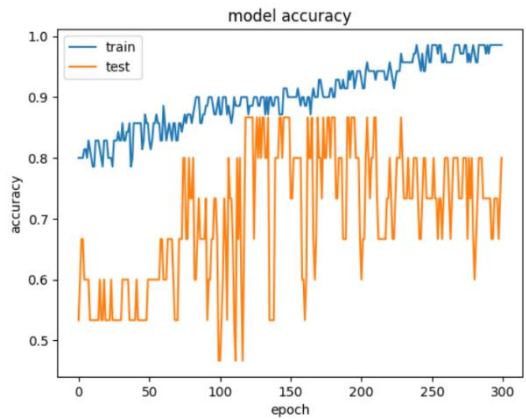

Figure 3: Training versus Validation Accuracy

Here TN, TP, FP, FN denote the true negative, true positive, false positive and false negative predictions respectively.

|  | **Accuracy** | **Loss** |
| --- | --- | --- |
| **Training** | 98.57 | 0.119 |
| **Validation** | 80.01 | 0.475 |

*4.1.2. Binary Cross-Entropy Loss*

Since there are two output categories in the dataset (likes denoted by 1 and dislikes denoted by 0), the most suitable function is binary cross-entropy loss. It quantifies the dissimilarity between true and predicted labels (Shannon, 1948). There is a penalty introduced every time an inaccurate prediction is made. It is calculated as follows:

$$Loss = \frac{1}{N}\sum_{i=1}^{N} -\left(y_i * log(p_i) + (1 - y_i) * log(1 - p_i)\right) \quad (6)$$

*4.2. Performance Analysis*

The training metrics indicate that the model has achieved adequate learning on the numerical parameters passed to it. The validation metrics also suggest that the model performs well on unseen data. However, the loss shows a lot of oscillation and plateaus, indicating saturation during testing. To address this, it may be necessary to add more complexity to the architecture. It is worthy of a note here that since the data is not time-series data, the model can only utilize dense layers.

The results demonstrate that the model has successfully learned intricate representations of the user's preferences and can accurately make predictions for songs that have not been listened to. However, it is important to consider that the model may not generalize well to new users.

# 5. Conclusion

Spotify has long been credited with commercializing and popularizing music recommendation systems well based on user preferences. There are two specific algorithms used in the same, the specifics of which are proprietary. Exploitative content is based on the user's previous history, and the previous history of similar users. This makes use of the already available data of a user to suggest more songs. Explorative or content-based filtering looks at the composition of the song in order to recommend a song. A balanced mix of the two strategies allow the user their "comfort zone" and encourage them to expand their taste gradually. For music recommendation, deep learning models are key to predict if a user is likely to enjoy a specific song based on their listening history and the audio features of the songs they have listened to. In context of content-based filtering, since the characteristics of the content, i.e. audio itself is used to predict recommendations, deep learning becomes especially useful to encapsulate patterns. A dataset of listening histories and respective ratings is often used with the aim of minimizing the difference between the predicted ratings and the actual ratings.

The model achieved significant accuracy in training and validation (98.57% and 80% respectively). These results in the aforementioned architecture indicate that the designed deep learning model effectively determined the likability of music for a user, achieving high accuracy in training and validation. This also implies that the model learned the complex representation of the user's preferences and predicted the likability of songs they had not listened to yet. It is worth noting that since the model is architecturally simple and takes into account only the sonic characteristics, there can be variations in results for a new user. Future work can include more comprehensive characteristics and a respective, depth-wise, and thorough architecture to encapsulate it.